# Análisis cuantitativo de riesgos utilizando "MCSimulRisk" como herramienta didáctica

Acebes F , Curto D , de Antón J , Villafáñez F

**Resumen**

La gestión del riesgo es una disciplina fundamental dentro de la gestión de proyectos, la cual incluye, entre otros, el análisis cuantitativo de los riesgos. A lo largo de varios años de docencia, hemos observado dificultades en los alumnos al realizar Simulación de Monte Carlo, dentro del análisis cuantitativo de los riesgos. El objetivo de este artículo es presentar "MCSimulRisk", como herramienta docente que permitirá a los estudiantes realizar simulación de Monte Carlo y aplicarlo a proyectos de cualquier complejidad, de una manera sencilla e intuitiva. Esta herramienta posibilita incorporar al modelo cualquier tipo de incertidumbre identificada en el proyecto.

**Palabras clave**

MCSimulRisk; Simulación de Monte Carlo; Análisis Cuantitativo de Riesgos; Gestión del Riesgo; Incertidumbre

## 1. Introducción

La inclusión de clases prácticas es esencial y complementaria a las clases teóricas. En estas sesiones, los estudiantes ponen en práctica los conocimientos adquiridos para resolver problemas prácticos, lo cual es fundamental para retroalimentar su aprendizaje. Al integrar adecuadamente estas clases prácticas en el desarrollo del curso, se refuerzan y aplican los conceptos teóricos, además de fomentar habilidades de análisis y síntesis en los estudiantes. Estas clases están estrechamente relacionadas con las clases teóricas, y de hecho, el profesor puede obtener información valiosa al detectar dificultades en la comprensión y aplicación de los conceptos. Es responsabilidad del profesor participar activamente para aclarar dudas o controversias que surjan, ya que en ese momento los estudiantes están más receptivos. Asimismo, el profesor debe observar cuidadosamente las respuestas de los estudiantes para evaluar la efectividad del programa educativo y las técnicas de enseñanza utilizadas en tiempo real (De Miguel Díaz, 2005).

La asignatura de Gestión del Riesgo, que se imparte en el Máster en Dirección de Proyectos de la Universidad de Valladolid, está centrada, principalmente, en el estudio de la incertidumbre del proyecto y en el análisis cuantitativo de los riesgos. Para poder trasmitir los conocimientos de una manera sólida a los alumnos, las clases teóricas se complementan con clases prácticas, donde los alumnos realizan ejercicios relacionados con la materia impartida, que les sirven para reforzar los conocimientos adquiridos.

El análisis cuantitativo de riesgos supone la realización de simulación de Monte Carlo, como método más adecuado para el tratamiento de la incertidumbre del proyecto, así como para la propuesta de indicadores de sensibilidad del proyecto. Existen en el mercado herramientas informáticas específicas para realizar esta actividad, como pueden ser Crystal Ball (Oracle, 2022a), Palisade - @Risk (Palisade, 2022), o incluso Oracle Primavera Risk Analysis (Pertmaster) (Oracle, 2022b). También existen otras herramientas más habituales con las que poder implementar simulación de Monte Carlo en proyectos. En este último caso nos referimos a simples hojas de cálculo (por ejemplo, Microsoft Excel (Microsoft, 2022)), donde se puede configurar el proyecto dado como problema y programar la simulación requerida.

En la asignatura que impartimos como docentes hemos utilizado algunas de las herramientas anteriormente

✉ Profesor Contratado Doctor Fernando Acebes Senovilla (1)
fernando.acebes@uva.es
ORCID: 0000-0002-4525-2610

Investigador Predoctoral David Curto Lorenzo (1)
david.curto@uva.es
ORCID: 0000-0002-4470-0004

Investigador Predoctoral Juan de Antón Heredero (1)
juan.anton@uva.es
ORCID: 0000-0002-2227-6505

Profesor Contratado Doctor Félix Villafáñez Cardeñoso (1)
felixantonio.villafanez@uva.es
ORCID: 0000-0001-7927-8627

(1) GIR INSISOC. Dpto. de Organización de Empresas y CIM. Escuela de Ingenierías Industriales. Universidad de Valladolid. Pº Prado de la Magdalena s/n, 47011 Valladolid (Spain).

mencionadas (Crystal Ball, @Risk y Excel). A pesar de que todas ellas han sido muy útiles para poder enseñar la asignatura a los alumnos y para que estos pudieran resolver los problemas planteados, nos hemos encontrado con dos problemas importantes. En primer lugar, tanto @Risk como Crystal Ball se ejecutan como un complemento (add-in) de Microsoft Excel. Por lo tanto, la programación del modelo de proyecto, las relaciones entre actividades, así como los resultados deseados, hay que programarlo bajo el entorno Excel. Esto supone una dificultad importante para los alumnos, especialmente cuando se trata de programar proyectos a partir de un determinado número de actividades, con un aumento de la complejidad de este y los alumnos únicamente tienen conocimientos básicos de Excel. El problema se manifiesta en el tiempo que el alumno emplea en la programación y modelización del proyecto y en la configuración de las actividades. Este tiempo dedicado a configurar el proyecto, podría ser utilizado de una manera más eficiente, en la comprensión del problema a resolver y en la discusión y razonamiento de los resultados que ofrece la simulación. Esto permitiría centrar el conocimiento del alumno en la materia (análisis cuantitativo del riesgo) y no tanto en la herramienta (Excel, @Risk, …).

El segundo de los inconvenientes detectados es que, a medida que aumenta la complejidad del proyecto, al pretender introducir más variables al proyecto en forma de riesgos estocásticos o epistémicos, la dificultad para realizar la programación del proyecto aumenta. Este problema nos conduce, inevitablemente, al problema anterior (pérdida irrecuperable de tiempo por parte de los alumnos).

Para resolver estos dos problemas mencionados con anterioridad, observados a lo largo de los últimos años de docencia en la asignatura, se decidió implementar la herramienta docente basada en Matlab, denominada "MCSimulRisk" (Monte Carlo Simulation for Risk Analysis). Esta herramienta nos va a permitir realizar simulación de Monte Carlo, para proyectos con cualquier grado de complejidad (cualquier número de actividades y cualquier relación de precedencia entre ellas) y, además, nos proporciona una forma sencilla de introducir en la simulación otros tipos de incertidumbre distintas de la incertidumbre aleatoria, como son la incertidumbre estocástica y la incertidumbre epistémica.

El objetivo de este artículo es presentar "MCSimulRisk" como la herramienta docente a utilizar en la asignatura Gestión del Riesgo, mostrando sus funcionalidades, sus características principales y los resultados que ofrece al usuario. Como objetivo secundario, pero no menos importante, se pretende que esta herramienta sea útil en actividades de seguimiento y control de proyectos en entornos con incertidumbre. Con la utilización de esta aplicación será posible obtener distintos indicadores relativos al control del proyecto en entornos con incertidumbre propuestos en la literatura, incluso incorporando al modelo incertidumbres de cualquier tipo identificadas en el proyecto. "MCSimulRisk" aborda fácilmente estas necesidades frente a la complejidad y dificultad que, a nuestro juicio, presentan el resto de las aplicaciones comerciales existentes.

El resto del documento se estructura como sigue. En el capítulo siguiente se realizará una revisión bibliográfica sobre la materia de estudio en la que se centra esta aplicación. Esto nos conduce a revisar los conceptos más importantes sobre la Gestión del Riesgo, la simulación de Monte Carlo, así como sobre otros conceptos asociados y a los que "MCSimulRisk" da respuesta. El capítulo 3 lo dedicamos a describir la aplicación. Para ello, utilizaremos un ejemplo de proyecto utilizado en clase, que nos servirá para describir los pasos a seguir, así como los resultados obtenidos, facilitando la comprensión del problema. Por último y para finalizar, exponemos las conclusiones extraídas de la realización de este trabajo.

## 2. Antecedentes

La gestión de riesgos en un proyecto abarca una serie de procesos que implican la planificación, identificación, análisis, planificación de respuesta, monitoreo y control de los riesgos. El propósito de la gestión de riesgos en un proyecto es incrementar la probabilidad y el impacto de eventos favorables, mientras que reduce la probabilidad y el impacto de eventos adversos para el proyecto (Project Management Institute, 2017).

Hoy en día, la Gestión de Riesgos se ha convertido en un componente indispensable de la Gestión de Proyectos, por lo que resulta inconcebible llevar a cabo la gestión de un proyecto sin llevar a cabo un análisis de los riesgos implicados. De este modo, cualquier organización que se embarque en un proyecto nuevo se encuentra con una serie de riesgos inherentes que necesitan ser abordados. En esta perspectiva, resulta fundamental establecer una política de riesgos que se difunda y aplique en todos los niveles de la organización, con el objetivo de mantener los riesgos dentro de los niveles aceptados y predefinidos. Al mismo tiempo, se busca prevenir o reducir los impactos de los riesgos negativos, mientras se busca maximizar los riesgos positivos.

La Dirección y Gestión de Proyectos cuenta con varias asociaciones profesionales a nivel mundial que han potencializado el desarrollo de la disciplina en el ámbito profesional y laboral. Estas asociaciones profesionales, enfocadas en el estudio de la gestión de proyectos, han creado, publicado y divulgado conocimientos especializados (BoKs) que sintetizan los avances más destacados y relevantes en este campo de conocimiento (Rozenes et al., 2006, 2004). Estos cuerpos de conocimiento incluyen los procesos, métodos, las herramientas y técnicas, y también las habilidades necesarias para el desarrollo de la dirección de proyectos (White and Fortune, 2002).

Los principales estándares, normas y metodologías en Dirección de Proyectos, como el PMBoK (Project Management Institute, 2017), Prince2 (Axelos, 2017), PM2 (European Commission, 2018), ICB 4.0 (International Project Management Association, 2015) o ISO 31000 (International Standards Organisation, 2018), dividen el proceso global de gestión del proyecto en diferentes áreas de conocimiento o procesos (cronograma, costes, calidad, etc). A su vez, cada uno de estos enfoques incluyen una fase o proceso específico para la Gestión del Riesgo (Tabla 1). Esta fase es abordada por los citados estándares y metodologías de forma diferente, utilizando estructuras alternativas, y con un grado de detalle de las herramientas y técnicas que difieren en los distintos procesos; no obstante, el contenido y la filosofía del proceso general de Gestión del Riesgo es común para todos ellos.

**Tabla 1** Fases o procesos de la Gestión del Riesgo según los distintos enfoques.

| PMBoK | PM² | Prince2 | ICB 4.0 | ISO 31000 |
|---|---|---|---|---|
| 1. Planificar la gestión de los riesgos | 1. Identificar los riesgos | 1. Identificar los riesgos | 1. Marco de gestión de riesgos | 1. Comunicación y consulta |
| 2. Identificar los riesgos | **2. Evaluar riesgos** | 1.1. Identificar el contexto | 2.Identificar oportunidades y amenazas | 2. Alcance, contexto, criterios |
| **3. Análisis cualitativo de los riesgos** | 3. Desarrollar respuesta a los riesgos | 1.2. Identificar los riesgos | **3. Evaluar riesgos** | **3. Evaluación del riesgo** |
| **4. Análisis cuantitativo de los riesgos** | 4. Control de los riesgos | **2. Evaluar los riesgos** | 4. Seleccionar la respuesta a los riesgos | **3.1. Identificación del riesgo** |
| 5. Planificar la respuesta a los riesgos | | **2.1. Estimación del riesgo** | 5. Controlar | **3.2. Análisis del riesgo** |
| 6. Implementar la respuesta a los riesgos | | **2.2. Evaluación del riesgo** | | **3.3. Valoración del riesgo** |
| 7. Monitorear | | 3. Planificar la respuesta a los riesgos | | 4. Tratamiento del riesgo |
| | | 4. Implementar respuesta a los riesgos | | 5. Seguimiento y revisión |
| | | 5. Comunicar | | 6. Registro e informe |

Analizando cada uno de los enfoques, podemos comprobar cómo, de forma explícita (PMBoK) o de forma implícita (PM2), en todos ellos se propone la realización de una evaluación cuantitativa de riesgos, dentro de la fase específica de evaluación del riesgo. El proceso de realizar el Análisis Cuantitativo de Riesgos consiste en analizar numéricamente el efecto de los riesgos identificados sobre los objetivos generales del proyecto.

Aunque existen distintas técnicas y herramientas que permiten realizar el análisis cuantitativo de riesgos (International Organization for Standardization, 2019), una de las estrategias más utilizadas es la simulación de Monte Carlo, que se emplea para incorporar una estructura de factores de riesgo y un modelo de dispersión de probabilidades para evaluar la incertidumbre (Kerzner, 2022). La simulación de Monte Carlo es una técnica que se utiliza para resolver problemas matemáticos mediante un modelo estadístico. Su enfoque radica en generar escenarios posibles a partir de datos iniciales, lo que permite analizar diferentes resultados potenciales. Este enfoque busca simular un escenario real y explorar sus diversas posibilidades, lo que permite al usuario predecir cómo se comportarán las variables con base en las estimaciones obtenidas a través de esta técnica (Kwak and Ingall, 2007).

En el campo de la dirección de proyectos, y más concretamente en el ámbito de la gestión del riesgo, la realización de simulación de Monte Carlo tiene muchas aplicaciones. Trataremos de resumir las más importantes, que coincidirán con aquellas que utilizamos en las clases de Gestión del Riesgo (Máster en Dirección de Proyectos), y que están implementadas en la herramienta educativa "MCSimulRisk".

En primer lugar, el análisis cuantitativo de riesgos tiene utilidad para el cálculo de las reservas de contingencias para el coste del proyecto (márgenes si hablamos del plazo del proyecto) (Kwon and Kang, 2019). Entendemos como reservas para contingencias al presupuesto añadido a la línea base de costes como consecuencia de los riesgos identificados (Project Management Institute, 2017). Se trata de una cantidad de fondos añadidos a la estimación de costes base para cubrir la incertidumbre estimada y la exposición al riesgo. A pesar de que existen una gran variedad de métodos, modelos y herramientas para el cálculo de contingencias (Islam et al., 2021), la simulación de Monte Carlo es uno de los métodos más utilizados y aceptados en la literatura (Barraza and Bueno, 2007; Liu et al., 2017; Vose, 2008). Como ejemplo destacado, Curto et al. (2022) utilizan simulación de Monte Carlo para el cálculo de contingencias de coste de un proyecto de construcción en el que se considera toda la información disponible sobre las incertidumbres identificadas en dicho proyecto.

Otra de las aplicaciones donde tradicionalmente se ha utilizado la simulación de Monte Carlo como herramienta para la gestión del riesgo, ha sido para el cálculo de indicadores de sensibilidad de las actividades del proyecto. Las métricas de sensibilidad de las actividades del proyecto miden la importancia de las actividades en el calendario de un proyecto (Acebes et al., 2021a; Vanhoucke, 2010). Las métricas más importantes utilizadas en dirección de proyectos (aunque no las únicas) son: Índice de Criticidad (CI), Índice de Crucialidad (CrI), y Schedule Sensitivity Index (SSI).

El Índice de Criticidad es un concepto introducido por Martin (1965) y representa el número de veces que una actividad pertenece al camino crítico del proyecto, considerando que las actividades que componen el proyecto tienen incertidumbre y variabilidad. Por su parte, el Índice de Crucialidad (Williams, 1992) mide la correlación que existe entre la duración de una actividad respecto de la duración total del proyecto. Este indicador complementa al anterior y su información debe ser manejada de forma compartida y no excluyente. Por último, el Schedule Sensitivity Index fue propuesto por el (Project Management Institute, 2004), y representa el Índice de Criticidad de la actividad corregido con la relación entre las desviaciones típicas de la duración de la actividad respecto de la duración total del proyecto.

Estas métricas de sensibilidad de las actividades son utilizadas en tareas de control del proyecto, para comprobar en cada instante de control si el proyecto se desvía respecto de los umbrales fijados previamente (Vanhoucke, 2011). Se ha programado en la aplicación "MCSimulRisk" la opción de cálculo y descarga de estos indicadores de sensibilidad correspondientes a las actividades del proyecto.

El análisis cuantitativo de riesgos y la simulación de Monte Carlo también se han utilizado en actividades de seguimiento y control del proyecto, en aquellos proyectos que presentan incertidumbre en la duración de sus actividades. Pasamos de un control determinista de los proyectos, utilizando Metodología del Valor Ganado (Anbari, 2003; Fleming and Koppleman, 2010; Lipke, 2003), a un control estocástico. Pajares & López-Paredes (2011) proponen la primera metodología que incorpora la variabilidad de las actividades para determinar el estado del proyecto. Los autores formulan la construcción de las curvas Schedule Risk Baseline (SRB) / Cost Risk Baseline (CRB) que representan la evolución del riesgo del proyecto (medido a través de la varianza), desde que el proyecto comienza hasta que finaliza. Utilizan estas curvas para proponer los indicadores Schedule Control Index (SCoI) / Cost Control Index (CCoI) con los que realizar el control del proyecto. Aprovechando la aparición de las curvas SRB/CRB, Acebes, Pajares, et al. (2021) proponen el indicador de sensibilidad Activity Risk Index (ARI) que mide la importancia de las actividades del proyecto en función del riesgo que aportan al mismo.

Más tarde surgen otras metodologías estocásticas, utilizando herramientas diferentes, con el objetivo de poder controlar el proyecto para mantenerlo dentro de los objetivos de plazo y coste establecidos (Colin et al., 2015; Wauters and Vanhoucke, 2014). Queremos resaltar las metodologías propuestas por Acebes et al. (2014, 2015) ya que serán incorporadas en "MCSimulRisk" como opciones para el usuario, para poder realizar el control estocástico del proyecto. En la primera de ellas (Acebes et al., 2014), se propone la metodología denominada "Triad", que incorpora unos indicadores que reflejan el estado del proyecto bajo control (retraso/sobrecoste), así como el rango porcentual en el que se encuentra en el momento de ejecución. En la segunda metodología (Acebes et al., 2015), los autores proponen incorporar técnicas estadísticas para pronosticar el tiempo y el coste a la finalización del proyecto. "MCSimulRisk" permitirá obtener los indicadores propuestos por Acebes et al. (2014, 2015) para poder realizar el control de aquellos proyectos en entornos inciertos, incluso considerando la incorporación de incertidumbres estocásticas o epistémicas.

## 3. Monte Carlo Simulation for Risk Analysis (MCSimulRisk)

"MCSimulRisk" (Monte Carlo Simulation for Risk Analysis) es una aplicación desarrollada en el entorno Matlab®, configurada para realizar simulación de Monte Carlo y aplicarlo al análisis cuantitativo de riesgos en los proyectos. El objetivo de "MCSimulRisk" es servir de herramienta docente para su utilización, por parte de los alumnos, en los laboratorios de la asignatura Gestión del Riesgo, del Máster en Dirección de Proyectos de la Universidad de Valladolid. Poner a disposición de los alumnos esta herramienta, permitirá que estos centren sus esfuerzos en la resolución del problema planteado, y utilicen el tiempo para la realización efectiva de las prácticas, evitando pérdidas de tiempo y de recursos en la programación del modelo (normalmente en la aplicación MS Excel). En la Figura 1 mostramos el diagrama de flujo de la aplicación con la secuencia de pasos que debe seguir el alumno durante su utilización, así como con las utilidades que ofrece a los usuarios.

**Figura 1** Diagrama de flujo aplicación "MCSimulRisk".

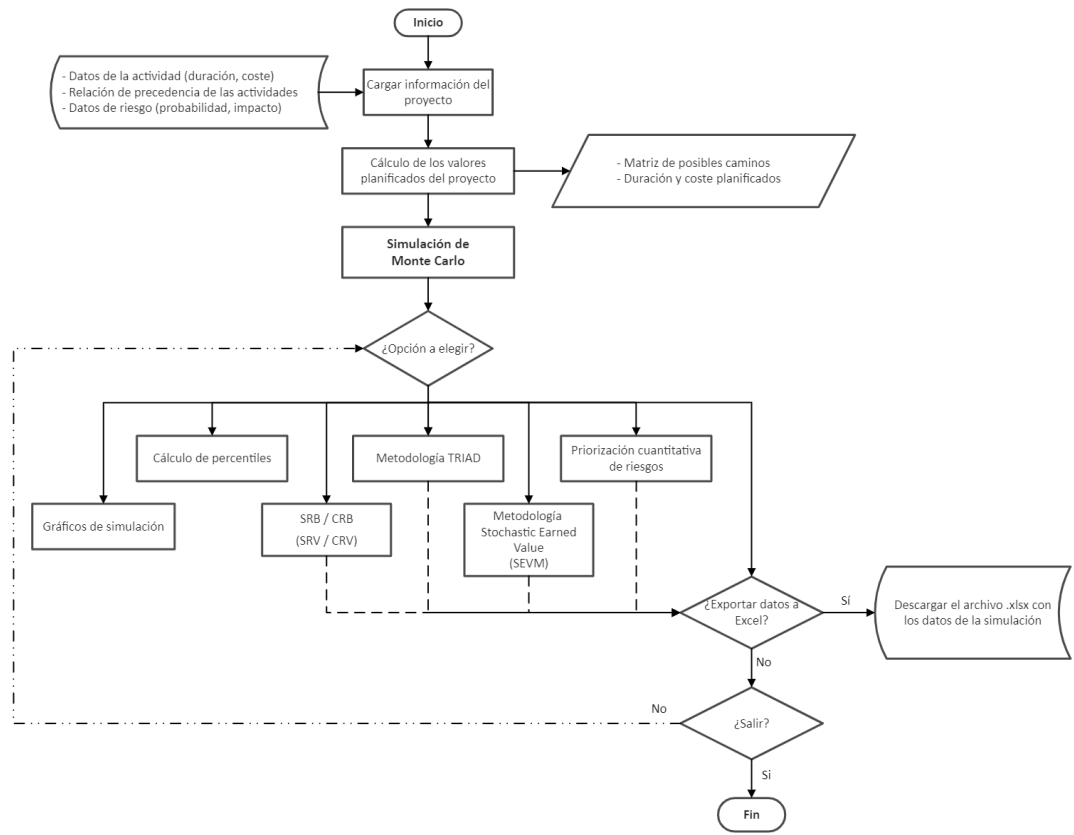

## 3.1. Definición de actividades

El proceso comienza con la definición de las actividades del proyecto. Para ello se utiliza una hoja Excel auxiliar, donde se identifica el número de actividades que componen el proyecto (Figura 2). En entornos estocásticos, la duración de las actividades está caracterizada por presentar incertidumbre aleatoria, por lo que habrá que definir la función de distribución que representa el comportamiento de la actividad, pudiendo seleccionar entre varias opciones según se haya definido el comportamiento aleatorio de la actividad (normal, uniforme, triangular, …). Además de la duración de cada actividad, se debe definir el coste de esta, diferenciando el "coste fijo", independiente de la duración de la actividad (coste relativo a materiales, subcontrataciones, equipos, …), y el "coste variable", relacionado directamente con la duración de la actividad (coste en mano de obra variable, por ejemplo).

A parte de la incertidumbre aleatoria con la que definimos la duración de las actividades, esta aplicación permite incorporar otros tipos de incertidumbre (o riesgos) al modelo de proyecto, diferenciando si son riesgos que impactan sobre la duración de alguna actividad de proyecto, o son riesgos que impactan sobre el coste de la actividad. Para este tipo de riesgos, debemos definir la probabilidad de ocurrencia de dicho riesgo, así como el posible impacto si estos llegasen a producirse. La aplicación incorpora distintos modelos de funciones de distribución a los que asignar tanto la probabilidad como el impacto del riesgo.

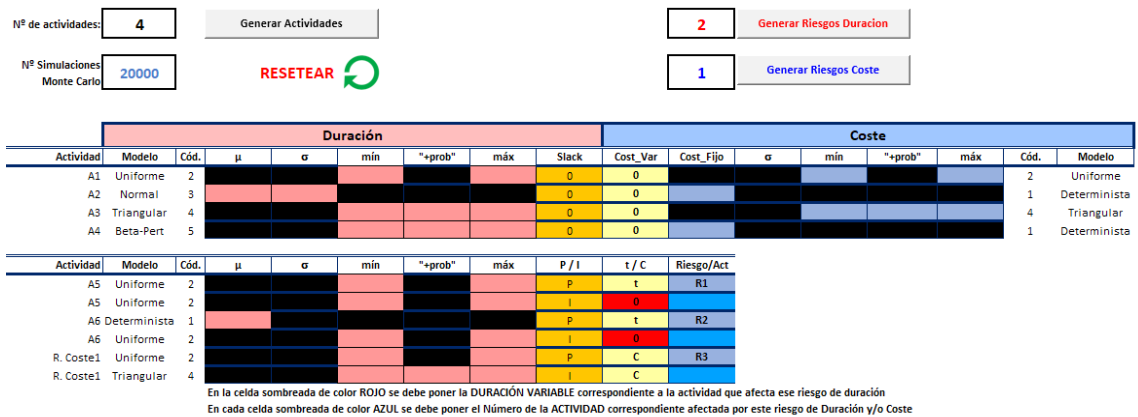

**Figura 2** Diagrama de flujo aplicación "MCSimulRisk".

Otro aspecto crucial que hay que definir previamente para poder operar con "MCSimulRisk" es la relación de precedencias entre actividades del proyecto. El calendario debe seguir las prácticas recomendadas de programación CPM (Critical Path Method) (AACE - American Association of Cost Engineering, 2011) porque permite calcular los caminos críticos con precisión como base para la simulación de Monte Carlo. La forma de incorporar esta información en la aplicación es a través de una matriz en la cual, un "1" en la celda de la matriz, significa que la actividad de la fila correspondiente a esa marca tiene como actividad predecesora a aquella actividad que identifica dicha columna (Figura 3). La matriz de relación de precedencias incluye los riesgos identificados que pueden impactar sobre la duración del proyecto y deberán modelarse como actividad sucesora de aquella sobre la que impacta dicho riesgo.

**Figura 3** Relación de precedencias de las actividades del proyecto.

| Precedentes | A0 | A1 | A2 | A3 | A4 | R1 A5 | R2 A6 | Af |
|---|---|---|---|---|---|---|---|---|
| A0 | 0 | 0 | 0 | 0 | 0 | 0 | 0 | 0 |
| A1 | 1 | | | | | | | |
| A2 | 1 | | | | | | | |
| A3 | | | | | | 1 | | |
| A4 | | | | | | 1 | 1 | |
| R1 A5 | | | 1 | | | | | |
| R2 A6 | | | | 1 | | | | |
| Af | | | | | 1 | 1 | | |

Durante la realización de las prácticas de laboratorio, el alumno deberá cumplimentar la tabla con toda la información de las actividades y riesgos (Figura 2), así como cumplimentar la matriz de relación de precedencias de las actividades, que incluye los riesgos con impacto en la duración del proyecto (Figura 3). Por lo tanto, el alumno no necesita tener conocimientos avanzados sobre programación en Excel para poder realizar los ejercicios de laboratorio. Conseguimos alcanzar uno de los objetivos planteados que consistía en que el alumno debe dedicar el tiempo útil de sus clases de laboratorio en estudiar el problema, encontrar soluciones y discutir resultados, en vez de utilizar su tiempo en modelar el proyecto objeto del problema, por desconocimiento de la herramienta Excel.

### 3.2. Resultados de la aplicación

Una vez se ha cargado la información del proyecto (actividades, precedencias y riesgos), ejecutamos la aplicación y "MCSimulRisk" nos ofrece los resultados deterministas del proyecto: valor esperado de las actividades, duración y coste planificados y matriz de posibles caminos del proyecto. A partir de ahí, la aplicación realiza simulación de Monte Carlo y disponemos de un menú con un listado de opciones a elegir en función del problema que se quiera resolver (Figura 4).

**Figura 4** a) Menú principal 'MCSimulRisk'. b) Menú representaciones gráficas 'MCSimulRisk'

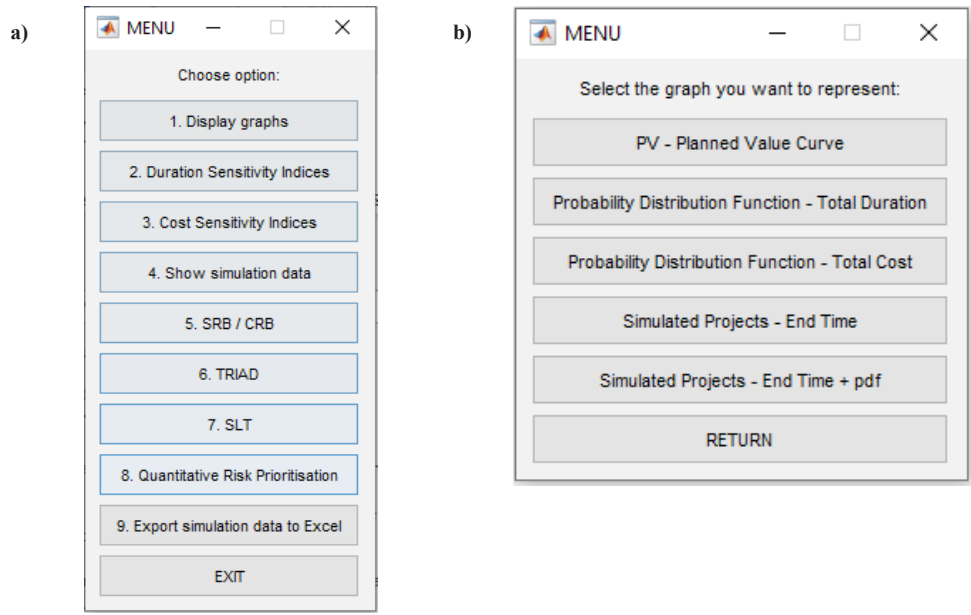

Como primera opción de las seleccionables en el menú aparece la representación gráfica de los resultados de la simulación (Figura 4 b): Curva de Valor Planificado, funciones de distribuciones de probabilidad (de duración y coste totales), que se representan conjuntamente con la función de distribución de probabilidad acumulada (Figura 5 a), y diagrama de dispersión de cada simulación en el instante final del proyecto (Figura 5 b). En la Figura 5 b) se representa la situación final de cada proyecto simulado en un gráfico cartesiano con ejes x-duración e y-costes. A la vez, se muestran las funciones de distribución de probabilidad para la simulación en cada eje correspondiente.

**Figura 5** a) Curvas de distribución de probabilidad y probabilidad acumulada correspondientes al coste total del proyecto. b) Diagrama de dispersión duración – coste, situación final del proyecto

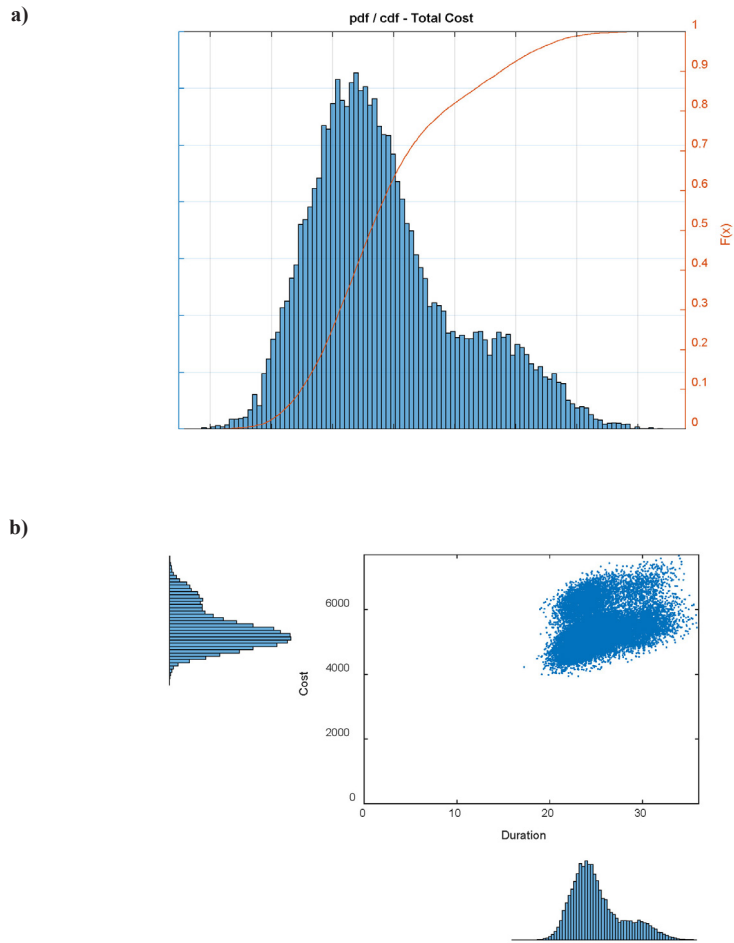

Además de conseguir las gráficas correspondientes a la duración y coste total del proyecto, "MCSimulRisk" proporciona los valores numéricos de percentiles de probabilidad, tanto de duración como de coste. Estos datos se pueden conseguir accediendo al menú principal y seleccionando la opción "Show simulation data".

Tal y como hemos mencionado en el apartado "Antecedentes", una de las aplicaciones más importantes de la simulación de Monte Carlo es el cálculo de indicadores de sensibilidad de duración de las actividades. Accediendo a la opción "Duration Sensitivity Index" del menú principal, podemos calcular, representar y exportar los datos a un fichero Excel (.xlsx), los principales índices de sensibilidad de las actividades del proyecto que incorpora "MCSimulRisk" (Figura 6).

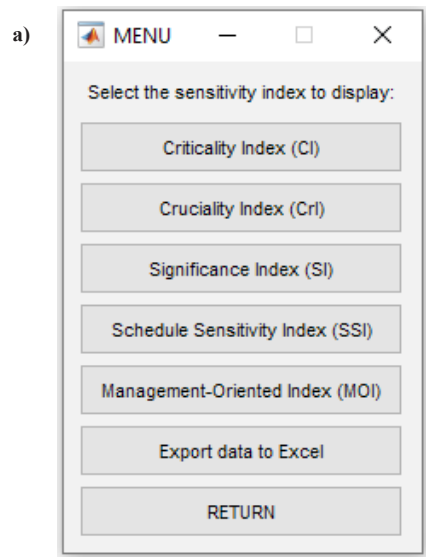 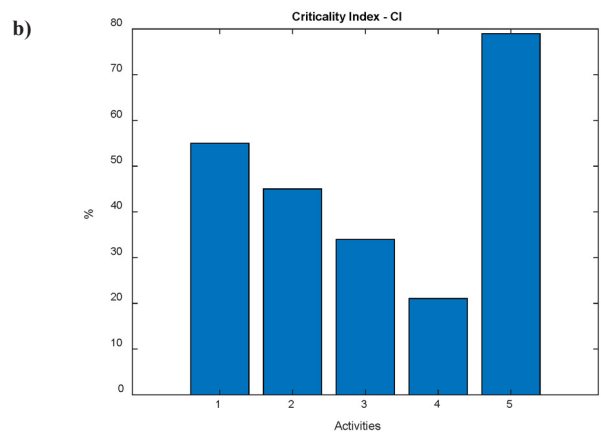

**Figura 6** a) Menú Índices de sensibilidad de las actividades. b) Representación gráfica Índice de Criticidad.

De nuevo comprobamos la utilidad de la herramienta presentada pues facilita al alumno la obtención de los resultados buscados, optimizando el tiempo de clase, sin destinarlo a otros usos distintos del objetivo de la asignatura (por ejemplo, utilizar el tiempo en programar en Excel las ecuaciones con las que calcular cada uno de los indicadores).

Continuando con las posibles aplicaciones que ofrece "MCSimulRisk", nos centramos en aquellas que nos permitirán realizar el control del proyecto en entornos con incertidumbre. Seleccionando la opción "SRB/CRB" del menú principal podremos calcular y representar las líneas base de riesgos SRB/CRB de programación y coste del proyecto (Pajares and López-Paredes, 2011) (Figura 7 a). La obtención de estas curvas son el paso previo para conseguir los indicadores SCoI/CCoI que proponen Pajares & López-Paredes (2011) y que se utilizan para conocer el estado del proyecto en el instante de control, conforme a la incertidumbre aportada por los riesgos identificados del proyecto (Acebes et al., 2013). Estudios posteriores realizados por Acebes, Poza, et al. (2021) utilizan las curvas anteriores para proponer el indicador Schedule Risk Value (SRV), con el que medir el riesgo total del proyecto, y el indicador Activity Risk Index (ARI), para priorizar la importancia de las actividades en función del riesgo que aportan al conjunto del proyecto.

"MCSimulRisk" permite utilizar la metodología Triad, propuesta por Acebes et al. (2014) (Figura 7 b). Con ella podremos realizar el seguimiento y control de proyectos en entornos de incertidumbre. En cada instante de control, el alumno podrá conocer en qué estado de sobrecoste o infracoste (retraso o adelanto, en el caso del tiempo) se encuentra su proyecto. Además, comparando el estado actual del proyecto en ejecución que el alumno quiere controlar, podrá conocer en dicho instante cómo se comporta respecto a las distintas curvas de probabilidad del proyecto planificado, es decir, en qué percentil se encuentra el proyecto en ejecución respecto de su planificación (tanto para el control del coste del proyecto como para la programación).

**Figura 7** a) Representación gráfica SRB – Metodología SCoI/CCoI. b) Representación gráfica Percentiles Programación - Metodología Triad.

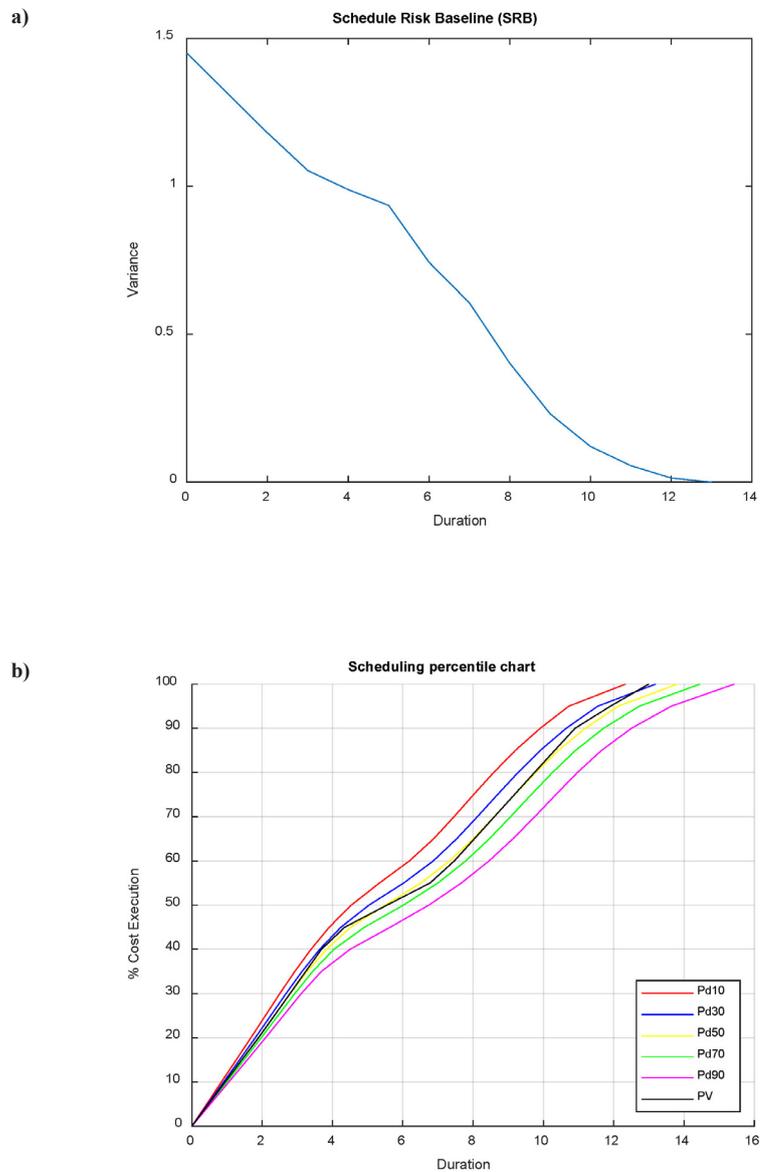

Por último, en cuanto a metodologías para el seguimiento y control de proyectos, "MCSimulRisk" nos proporciona la información necesaria para implementar la metodología Stochactic Earned Value Methodology (SEVM) (Acebes et al., 2015) (Figura 8). Esta metodología utiliza como dato de partida el porcentaje de ejecución del proyecto en el instante de control. Conocido este valor, la metodología emplea técnicas estadísticas para estimar la duración y costes finales del proyecto a partir de la situación (duración/coste) de cada uno de los proyectos simulados, tanto en el instante final del proyecto como en el instante que se corresponde con el porcentaje de ejecución en el que se encuentra el proyecto.

**Figura 8** a) Diagrama de dispersión Duración/Coste. Proyectos simulados en instante de control y en instante final – Metodología SEVM. b) Clasificación retrasos – Metodología SEVM.

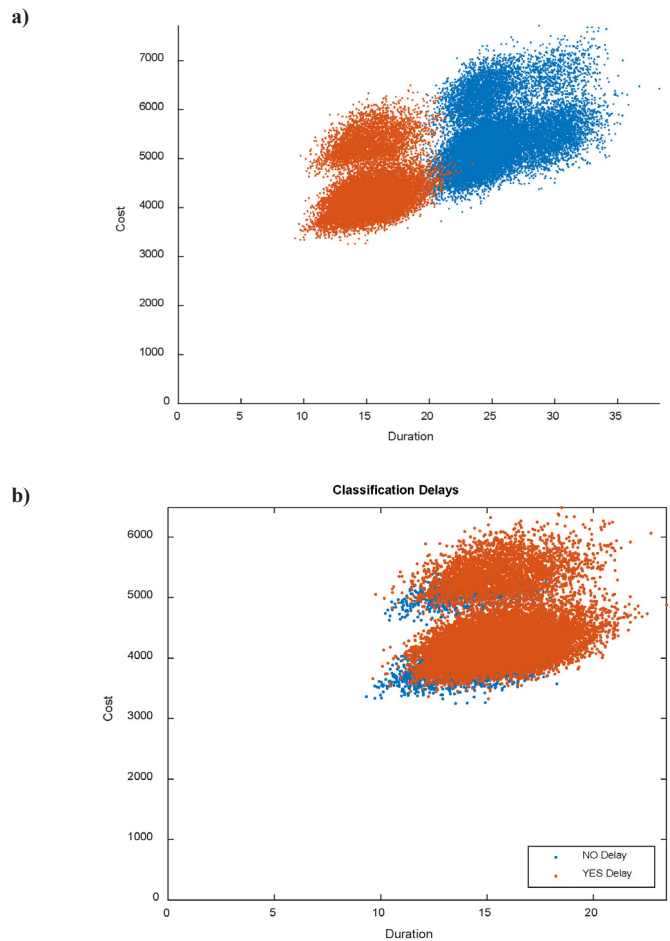

La aplicación docente "MCSimulRisk" facilita todos los datos necesarios al alumno para que pueda pronosticar si el proyecto real retrasará o adelantará (tendrá sobrecostes o ahorros) respecto de la planificación. Para ello, algunas de las gráficas facilitadas por la aplicación, y que el alumno puede utilizar, son el diagrama de dispersión que incluye los proyectos simulados, tanto en el instante final de ejecución como en el instante correspondiente al tiempo de control (Figura 8 a); o la gráfica que representa el conjunto de proyectos simulados en el instante de control del proyecto real (Figura 8 b). En este último caso, se han diferenciado los proyectos según será su comportamiento final con respecto a la duración planificada. En esta Figura 8 b, si los proyectos simulados van a finalizar con un retraso sobre el cronograma planificado se representan con color rojo, mientras que, si los proyectos van a finalizar en adelanto respecto del proyecto planificado, se colorean en azul.

## 4. Conclusiones

Se trata de dar solución a los problemas detectados con los que se enfrentan los alumnos durante la realización de los ejercicios de laboratorio de la asignatura Gestión del Riesgo, en el Máster en Dirección de Proyectos de la Universidad de Valladolid. Para ello, se ha desarrollado la aplicación "MCSimulRisk". Esta aplicación docente, desarrollada en el entorno Matlab, permite realizar simulación de Monte Carlo, con aplicación directa al análisis cuantitativo de riesgos de los proyectos.

En primer lugar, los alumnos que utilizan "MCSimulRisk" ya no necesitan tener conocimientos avanzados de programación en Excel. Es posible la configuración de cualquier tipo de proyecto, a pesar de la complejidad de su red (alto número de actividades o complejas redes serie/paralelo). Los alumnos únicamente tienen que definir las actividades que forman parte del proyecto, y de los riesgos que pueden impactar sobre él, para poder obtener los resultados deseados, sin utilizar la mayor parte del tiempo útil del laboratorio en configurar la red de proyecto y programar los resultados en la hoja Excel. "MCSimulRisk" ofrece todos los resultados necesarios que el alumno va a necesitar para la realización de sus prácticas (resultados gráficos, resultados numéricos y posibilidad de exportar los datos en un fichero externo para un tratamiento posterior).

En segundo lugar, existe la posibilidad de incorporar cualquier tipo de incertidumbre en la red de proyecto. Si tradicionalmente se ha tratado únicamente con la incertidumbre aleatoria de las actividades, "MCSimulRisk" permite incorporar los riesgos estocásticos y epistémicos identificados en el proyecto, diferenciando entre aquellos que pueden impactar sobre la duración del proyecto o los que pueden impactar sobre el coste. El coste de tiempo que debe emplear el alumno para realizar la incorporación de riesgos al modelo es mínimo; es suficiente con incluirlos en la identificación de riesgos previa a realizar la simulación de Monte Carlo.

Además, "MCSimulRisk" incorpora herramientas para realizar el seguimiento y control de los proyectos en entornos con incertidumbre. No sólo aporta resultados sobre indicadores de sensibilidad de las actividades del proyecto, o los resultados de la simulación (en duración y coste), sino que, además, permite utilizar distintas metodologías que tratan el control del proyecto con incertidumbre. "MCSimulRisk" ayudará a dar a conocer estas metodologías y a su utilización por parte de la comunidad científica, profesional y educativa. La ausencia de herramientas como la propuesta en este artículo, que traten la incertidumbre de una manera sencilla, provoca el rechazo a la implementación real de estas metodologías más actuales.

Por último, y no menos importante, "MCSimulRisk" abre un abanico importante de posibilidades para los investigadores, pues disponen de una herramienta que permite analizar el comportamiento de los riesgos estocásticos y epistémicos en la red del proyecto. Es de esperar que los resultados que puedan obtenerse al considerar todos los riesgos identificados en el proyecto sean diferentes de aquellos resultados que se obtienen al considerar únicamente los riesgos aleatorios de las actividades. Hasta la fecha, la literatura se ha centrado únicamente en incorporar estos últimos tipos de riesgos en los modelos de proyecto (aleatorios), pero con la utilización de "MCSimulRisk" esperamos avanzar en la investigación proporcionando una nueva visión sobre el comportamiento de todos los riesgos sobre el proyecto en su conjunto.

## Financiación

## Bibliografía